\documentclass[12pt]{article}

\usepackage[left=1in,right=1in,top=1in,bottom=1in]{geometry}
\usepackage{graphicx}
\usepackage{natbib}
\defcitealias{carb_2022}{CARB, 2022}
\defcitealias{cea_2024}{CEA, 2024}
\defcitealias{kern_2024}{Kern, 2024}
\usepackage{setspace}
\usepackage{soul}               
\usepackage{caption}
\usepackage{booktabs}
\usepackage{threeparttable}
\usepackage{siunitx}            
\usepackage{scalerel}
\usepackage{nicefrac}

\usepackage[]{appendix}

\usepackage[xcolor=dvipsnames]{changes}

\usepackage[shortlabels,inline]{enumitem}

\setlist[description]{leftmargin=0em,labelindent=\parindent}

\usepackage{amsmath,amsthm}
\usepackage{mathtools}

\theoremstyle{definition}

\theoremstyle{remark}

\usepackage[T1]{fontenc}

\usepackage[scaled=.85]{beramono}
\usepackage[type1]{cabin}
\usepackage{amsmath,amsthm}
\usepackage[libertine]{newtxmath}

\usepackage[scr=rsfso]{mathalfa}
\usepackage{bm}
\usepackage[semibold]{libertine}

\usepackage{titlesec}

\titleformat{\section}[block]{\bf\Large}{\thesection\quad}{0pt}{}
\titleformat{\subsection}[block]{\bf\large}{\thesubsection\quad}{0pt}{}

\newenvironment{figurenotes}[1][Note]{\begin{minipage}[t]{\linewidth}\footnotesize{\itshape#1: }}{\end{minipage}}

\usepackage[dvipsnames]{xcolor}
\definecolor{csub-blue}{RGB}{0, 53, 148}
\definecolor{csub-blue-hl}{RGB}{0, 53, 148}
\usepackage[unicode,hyperfootnotes=true,linktocpage,pagebackref]{hyperref}
 \hypersetup{%
   colorlinks = true,
   pdfborder = {0 0 0},
   citecolor = csub-blue,
   linkcolor = csub-blue,
   urlcolor  = csub-blue,
 }%
\makeatletter
\def\@fnsymbol#1{\ensuremath{\ifcase#1\or *\or \mathsection\or \mathparagraph\or \|\or **\or \dagger\dagger \or \ddagger\ddagger \else\@ctrerr\fi}}
\makeatother

\title{\textsf{\Large Not All Oil Price Shocks Are Alike.\\
    A Replication of Kilian (\textit{American Economic Review}, 2009)}.\thanks{We gratefully acknowledge support from the Resilient Energy Economies Initiative.
    Replication materials are available at
  \href{https://github.com/richryan/jcre-oil}{\nolinkurl{https://github.com/richryan/jcre-oil}}.}}
\author{Rich Ryan\thanks{Email: \href{mailto:richryan@csub.edu}{richryan@csub.edu}. Department of Economics, California State University, Bakersfield.}
  \and
  Nyakundi Michieka\thanks{Email: \href{mailto:nmichieka@csub.edu}{nmichieka@csub.edu}.
    Department of Economics, California State University, Bakersfield.}} 
\date{June 19, 2025}

\newcommand{\dateOrigStart}{February 1973}
\newcommand{\dateOrigEnd}{December 2007}
\newcommand{\dateUpdateStart}{February 1974}
\newcommand{\dateUpdateEnd}{December 2007}
\newcommand{\dateFullStart}{February 1974}
\newcommand{\dateFullEnd}{January 2025}
\newcommand{\dateFullWIPStart}{February 1974}
\newcommand{\dateFullWIPEnd}{January 2025}

\newcommand{\dateKernStart}{January 1990}
\newcommand{\dateKernEnd}{January 2025}

\begin{document}
\maketitle

\begin{abstract}
  The price of oil can rise because of a disruption to supply or an increase in demand.
  The nature of the price change determines the dynamic effects.
As \citet{kilian_2009} put it:
``not all oil price shocks are alike.''
Using the latest available data,
we extend \citeauthor{kilian_2009}'s analysis using the R ecosystem and provide more evidence for \citeauthor{kilian_2009}'s conclusions.
Inference based on unknown conditional heteroskedasticity
strengthens the conclusions.
With the updated shocks,
we assess how
a local economy responds to the global oil market,
an application that is relevant to policymakers
concerned with the transition away from fossil fuels.
\end{abstract}

\vspace{3em}


\vspace{1em}


\vfill

\clearpage
\pagebreak

\section{Introduction}

In a stylized model for the oil market,
the price of oil depends on supply and demand.
Price can go up if 
demand shifts outward or
supply shifts inward.
Even though the nature of the shocks differs, each causes the price of oil to rise.
Surprisingly, as \citet{kilian_2009} points out,
much theoretical, empirical, and policy-oriented work
focuses on the effects of oil-price fluctuations,
ignoring the nature of shocks.
This risks flawed policy.
Fluctuations in the price of oil are an amalgam of demand and supply shocks.
The dynamics generated by a demand shock differ from
the dynamics generated by a supply shock.
At any point in time,
the composition of shocks may be far from the average,
making correlations between the price of oil and US macroeconomic aggregates
potentially meaningless and uninformative for policymakers \citep{kilian_2008,kilian_2009}.

In brief, here is the main issue:
\citet{kilian_2009} establishes that the dynamic effects of an increase in the price of oil
depend on the nature of the increase---whether
the increase was the result of a
disruption to supply or an increase in demand.
The conclusions are based on a structural model of the global market for oil
estimated over the period \dateOrigStart{} to \dateOrigEnd.
Do the same conclusions hold using all available data?
Are the results robust to inference that
allows for conditional heteroskedasticity?

Using 
data on oil production and price 
from the US Energy Information Administration
and
data on global real economic activity available through FRED,
we estimate the structural vector autoregressive model of the global crude-oil market proposed by \citet{kilian_2009}.
The updated results provide further evidence that \citeauthor{kilian_2009}'s \citeyearpar{kilian_2009} conclusions hold.
In addition,
a procedure that accounts for
conditional heteroskedasticity of unknown form in the VAR model,
indicates that Kilian's conclusions are even stronger
than initially stated \citep{bruggemann_jentsch_trenkler_2016,kilian_lutkepohl_2017}.

The structural VAR model disentangles supply shocks from demand shocks in the global crude-oil market.
In the model,
supply comprises global oil production and
demand comprises two components:
\begin{enumerate*}[(i)]
\item\label{item:1} a component that reflects the global business cycle and 
\item a component that reflects oil-specific demand
  \setcitestyle{square}
  (buyers concerned about availability in the future will buy out of precaution, for example \citep{alquist_kilian_2010}). \setcitestyle{round}
\end{enumerate*}
This allows \citet{kilian_2009}
to show how each shock separately determines fluctuations in the real price of crude oil.
Moreover,
how real Gross Domestic Product responds to an unanticipated increase in the price of oil depends on the nature of the shock.
The same conclusion applies to the Consumer Price Index.
\citeauthor{kilian_2009}'s results explain the macroeconomic implications of oil-price fluctuations.

In a different direction,
we use the updated series of structural shocks to learn about another important policy question:
how does a local labor market
respond to these shocks? 
The question is of interest to policymakers in California.
California has adopted the goal of carbon neutrality by 2045 or earlier \citepalias{carb_2022}.
In Kern County,
which accounts for 70 percent of California's oil production,
there is concern about potential job loss and lost tax revenue as the economy transitions away from fossil fuels.
Yet,
given the complex linkages between oil extraction and the rest of the economy,
Kern's dependence on oil is hard to assess \citep[][describe one complex link]{golding_kilian_2022}.
We document how Kern's labor market responds to identified shocks in the global market for oil.
And we find that the response depends on the nature of the shock.
Kern's dependence on oil suggests that economic forces will restructure the composition of regional economic activity
along the path to net-zero emissions by 2045.
Past structural changes, like the transition away from coal, do not augur well for displaced workers and regions \citepalias{cea_2024}.
The evidence suggests that
policymakers may need to consider
expanding options in Kern through, for example, place-based policy.\footnote{What will happen in California
  as 2045 approaches is impossible to know with certainty.
  In particular, a backward-looking structural VAR model will not be a good predictor of abrupt change.
  Instead, we view Kern's responses to identified structural innovations as suggestive evidence that the Kern economy depends on oil.}
The exercise shows how \citeauthor{kilian_2009}'s \citeyearpar{kilian_2009} pioneering methods and conclusions apply more generally.

Questions about how oil affects the US economy are not fully answered.
Further research will benefit from other contributions of our replication.
We port Kilian's original MATLAB code to R.
Some researchers may prefer to work in R as
the R ecosystem makes tidying data easier.
(Our code embraces the \texttt{tidyverse}.)
We use the \texttt{targets} package to ensure replicability.
The \texttt{targets} package provides Make-like tools that
automatically make any necessary changes to dependencies when any change is registered.
In addition,
we retrieve data from the US Energy Information Administration and FRED using code,
which makes estimating the most recent structural shocks as straightforward as rerunning the code.
These resources
will make addressing questions easier in the ongoing debate about
the effects oil-price shocks have on the US economy.\footnote{An introduction to Make is
  provided by Jes{\'u}s Fern{\'a}ndez-Villaverde in \href{https://www.sas.upenn.edu/~jesusfv/teaching.html}{lecture notes}
  titled ``Computational Methods for Economists,'' which are available at \href{https://www.sas.upenn.edu/~jesusfv/teaching.html}{\nolinkurl{https://www.sas.upenn.edu/~jesusfv/teaching.html}}}

We present 
a replication of \citeauthor{kilian_2009}'s \citeyearpar{kilian_2009} results in section \ref{sec:rep}.
We conclude that \citet{kilian_2009} should be praised and thanked
for writing and sharing code
that makes reproducing results straightforward \citep{kilian_2009data}.
Using the posted data and our code,
we are able to exactly replicate \citeauthor{kilian_2009}'s \citeyearpar{kilian_2009} results.

The replication comprises
a description of the data (section \ref{sec:rep}),
a specification of the structural VAR model (section \ref{sec:svar}),
a presentation of the estimated impulse response functions (section \ref{sec:irfs}),
a decomposition of the real price of oil into identified components (section \ref{sec:hdecomp}), and
a second-stage analysis of how US real GDP and the Consumer Price Index respond to the identified structural shocks.
In section \ref{sec:update},
building upon an analysis undertaken by \citet{kilian_lutkepohl_2017},
we strengthen \citeauthor{kilian_2009}'s \citeyearpar{kilian_2009} conclusions by
estimating the structural VAR model using data from \dateFullStart{} to \dateFullEnd{} and
constructing confidence intervals for impulse responses that are robust to arbitrary forms of conditional heteroskedasticity.
To demonstrate the wider applicability of the results,
in section \ref{sec:app}, we analyze how the global oil market affects outcomes in a local labor market.

\section{Replication}
\label{sec:rep}

\citeauthor{kilian_2009}'s \citeyearpar{kilian_2009} pioneering model of the global crude-oil market specifies three determinants.
The methodology separates a single supply shock from two demand shocks. 
The oil-supply shock reflects unexpected disruptions to crude-oil availability.
The aggregate-demand shock reflects the consensus that shifts in global real economic activity shift current demand for crude oil \citep{kilian_zhou_2018}.
And the precautionary-demand shock reflects how demand may shift based on uncertainty about future unavailability of oil.
\citet{alquist_kilian_2010} make this point, among others, in an economy that receives a random endowment of oil each period.
They show that an increase in the risk of future availability of oil raises current demand. 

Three variables comprise
\citeauthor{kilian_2009}'s \citeyearpar{kilian_2009} vector autoregressive model of the global crude-oil market:
\begin{enumerate*}[(i)]
\item the log difference of global crude-oil production,
\item a measure of cyclical variation in global real economic activity, and
\item the log of the real price of oil.
\end{enumerate*}
Data on global crude-oil production are available from the \href{https://www.eia.gov/}{US Energy Information Administration} (EIA).
\citeauthor{kilian_2009}'s \citeyearpar{kilian_2009} measure of cyclical variation in global real economic activity
is made available by the Federal Reserve Bank of Dallas through the \href{https://fred.stlouisfed.org/}{FRED database}.
\citet{kilian_2019} reviews the index and corrects an error that was made when the original series was constructed.
The real price of oil is constructed by deflating US refiners' acquisition cost of imported crude oil (available from the EIA) by the
Consumer Price Index for All Urban Consumers.
Some details about the series and how we accessed them are included in an appendix in section \ref{sec:data}.\footnote{The appendix also reports a sensitivity analysis,
in which we replace \citeauthor{kilian_2009}'s \citeyearpar{kilian_2009} index with the monthly growth rate of
OECD+6 industrial production,
which is an index of industrial production for OECD (Organization for Economic Co-operation and Development) countries and
six non-member countries (Brazil, China, India, Indonesia, the Russian Federation, and South Africa).
Further discussion is provided in the appendix.}

\subsection{The structural VAR model}
\label{sec:svar}

To be precise about how
the determinants of the global market for crude oil are related,
consider a VAR model for 
$y_t = \left( \Delta \text{oil production}, \text{real activity}, \text{real price of oil}\right)^{\prime}$.
The structural VAR representation is
\begin{equation}
  B_0 y_t = \beta + B_1 y_{t-1} + \cdots + B_{24} y_{t-24} + \varepsilon_t,
  \label{eq:svar}
\end{equation}
where the $B_{i}$ are $3 \times 3$ parameter matrices and $\varepsilon_t \sim \left( 0, \Sigma_{\varepsilon}\right)$; that is,
$\varepsilon_t$ denotes a vector of serially and mutually uncorrelated innovations.
The covariance matrix $\Sigma_{\varepsilon}$ is diagonal.
The corresponding reduced-form model is
\begin{equation}
  y_t = \alpha + A_1 y_{t-1} + \cdots + A_{24} y_{t-24} + \epsilon_t,
  \label{eq:ols}
\end{equation}
where $A_i = B_0^{-1}B_i$ and $\epsilon_t = B_0^{-1} \varepsilon_t \sim \left( 0 , \Sigma_{\epsilon} \right)$
is a martingale-difference sequence with positive-definite covariance matrix
$\Sigma_{\epsilon} = \left( B_0^{-1} \right) \Sigma_{\varepsilon} \left( B_0^{-1} \right)^{\prime}$.
Going between the structural model in \eqref{eq:svar} and the reduced-form model in \eqref{eq:ols}
requires an estimate of $B_0^{-1}$.
Short-run identifying restrictions make estimating this term straightforward.

The recursive structure for $B_0^{-1}$ implies that the reduced-form error $\epsilon_t$ can be decomposed as
\begin{equation}
\epsilon_t 
\equiv 
\begin{bmatrix*}[l]
\epsilon_t^{\Delta \text{oil production}} \\
\epsilon_t^{\text{real activity}} \\
\epsilon_t^{\text{real price of oil}} 
\end{bmatrix*}
= 
\begin{bmatrix}
b_{11} & 0      & 0 \\
b_{21} & b_{22} & 0 \\
b_{31} & b_{32} & b_{33}
\end{bmatrix}
\begin{bmatrix*}[l]
\varepsilon_t^{\text{oil supply}} \\
\varepsilon_t^{\text{aggregate demand}} \\
\varepsilon_t^{\text{oil-specific demand}} 
\end{bmatrix*}.
\label{eq:identification}
\end{equation}
Within the system of three equations,
the first equation indicates that the stochastic short-run supply curve is vertical.
In other words, oil production does not respond to innovations in demand within the month.
The maintained assumption is that
decisions about oil supply are based on medium-run expectations and, 
because adjusting supply is costly,
short-run demand innovations are postulated to have a negligible affect on medium-run expectations.
Evidently, the model allows energy prices to respond to all past information.
There are two demand shocks and
each is identified by delay restrictions.
The short-run demand curve is downward sloping and can be shifted by 
innovations to aggregate demand and innovations to oil-specific demand,
such as precautionary-demand innovations \citep{alquist_kilian_2010}. 
The delay restriction is that oil-specific demand shocks do not affect
global real economic activity as measured by fluctuations in shipping rates relative to trend.
These postulations are further discussed by \citet{kilian_2008} and \citet{kilian_2009}.

In \eqref{eq:identification}, $B_0^{-1}$ is lower triangular.
A lower-triangular Cholesky decomposition can therefore be used
to decompose the estimate for $\widehat{\Sigma}_{\epsilon}$,
starting from the reduced-form estimates of the model in \eqref{eq:ols}.
Uncovering the structural shocks is possible using the estimated relationships in \eqref{eq:identification}.

Inference is based on the wild bootstrap discussed by \citet{goncalves_kilian_2004}
with 1,000 replications.
\citet{kilian_2009} uses 2,000 replications but the results are similar.
The wild bootstrap allows for conditional heteroskedasticity of unknown form.
But the asymptotic validity of the wild bootstrap is for the 
reduced-form impulse responses and not the asymptotic distribution of the error--covariance matrix.
This feature is 
shown by \citet{bruggemann_jentsch_trenkler_2016} and
discussed by \citet{kilian_lutkepohl_2017}.
In section \ref{sec:update},
we follow \citet{kilian_lutkepohl_2017} and adopt the inference procedures proposed by \citet{bruggemann_jentsch_trenkler_2016}
to investigate whether the larger confidence intervals matter in practice.

\subsection{How global oil production, real economic activity, and the real price of oil respond to demand and supply shocks in the crude-oil market}
\label{sec:irfs}

\begin{figure}[htbp]
\centerline{\includegraphics[width=\textwidth]{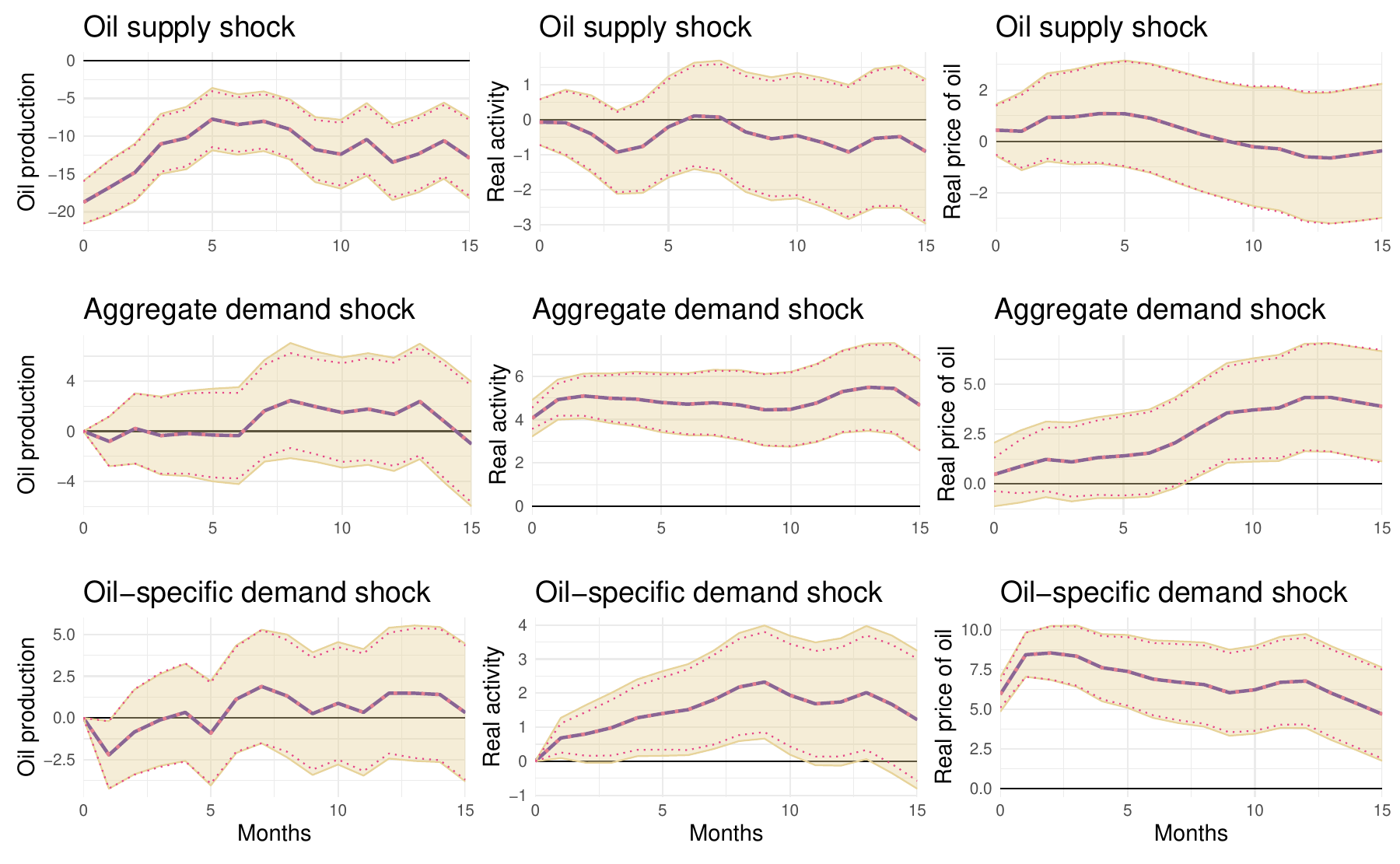}}
\caption[]{\label{fig:irf-replication} Responses to one-standard-deviation structural shocks (based on original data)}
\begin{figurenotes}[Note]
  Solid lines show the point estimates and the shaded regions show two-standard-error bands.
  Estimates are based on the structural VAR model in \eqref{eq:svar},
  using the original data \citep{kilian_2009data} (\dateOrigStart--\dateOrigEnd).
  Confidence intervals were constructed using a recursive-design wild bootstrap.
  Dotted lines show the point estimates and two-standard-error bands produced by \citeauthor{kilian_2009}'s \citeyearpar{kilian_2009} code.
  The striped effect indicates the two sets of point estimates coincide. 
\end{figurenotes}
\end{figure}

The structural impulse responses
define the responses of elements of $y_t$ to one-time impulses in $\varepsilon_t$.
Figure \ref{fig:irf-replication}
shows how global oil production, real economic activity, and the real price of oil
respond to one-standard deviation structural innovations.
The innovations are normalized so that each innovation will tend to raise the price of oil or,
put differently and slightly imprecisely,
figure \ref{fig:irf-replication} shows the effects of negative supply shocks and positive demand shocks.
Figure \ref{fig:irf-replication} is comparable to figure 3 in \citet{kilian_2009}.

Figure \ref{fig:irf-replication} uses data provided by \citet{kilian_2009data}.
The data start in
\dateOrigStart{}
and end in
\dateOrigEnd.
The goal of figure \ref{fig:irf-replication} is to assess whether the R code written for the replication
yields similar results to the MATLAB code that produces the original graphics.

Figure \ref{fig:irf-replication} confirms that the two sets of code produce identical results.
The solid, blue lines show the point estimates of the structural impulse responses.
They match those produced by the code provided by \citet{kilian_2009} exactly.
\citeauthor{kilian_2009}'s \citeyearpar{kilian_2009} original estimates are shown using red, dotted lines.
These overlay the solid, blue lines, which produces a striped effect.
The shaded areas depict the two-standard-error bands constructed by the wild bootstrap.
The dotted, red lines show those produced by the original code.
Again, there is overwhelming overlap,
which gives confidence that the two sets of codes are working.

\begin{figure}[htbp]
\centerline{\includegraphics[width=\textwidth]{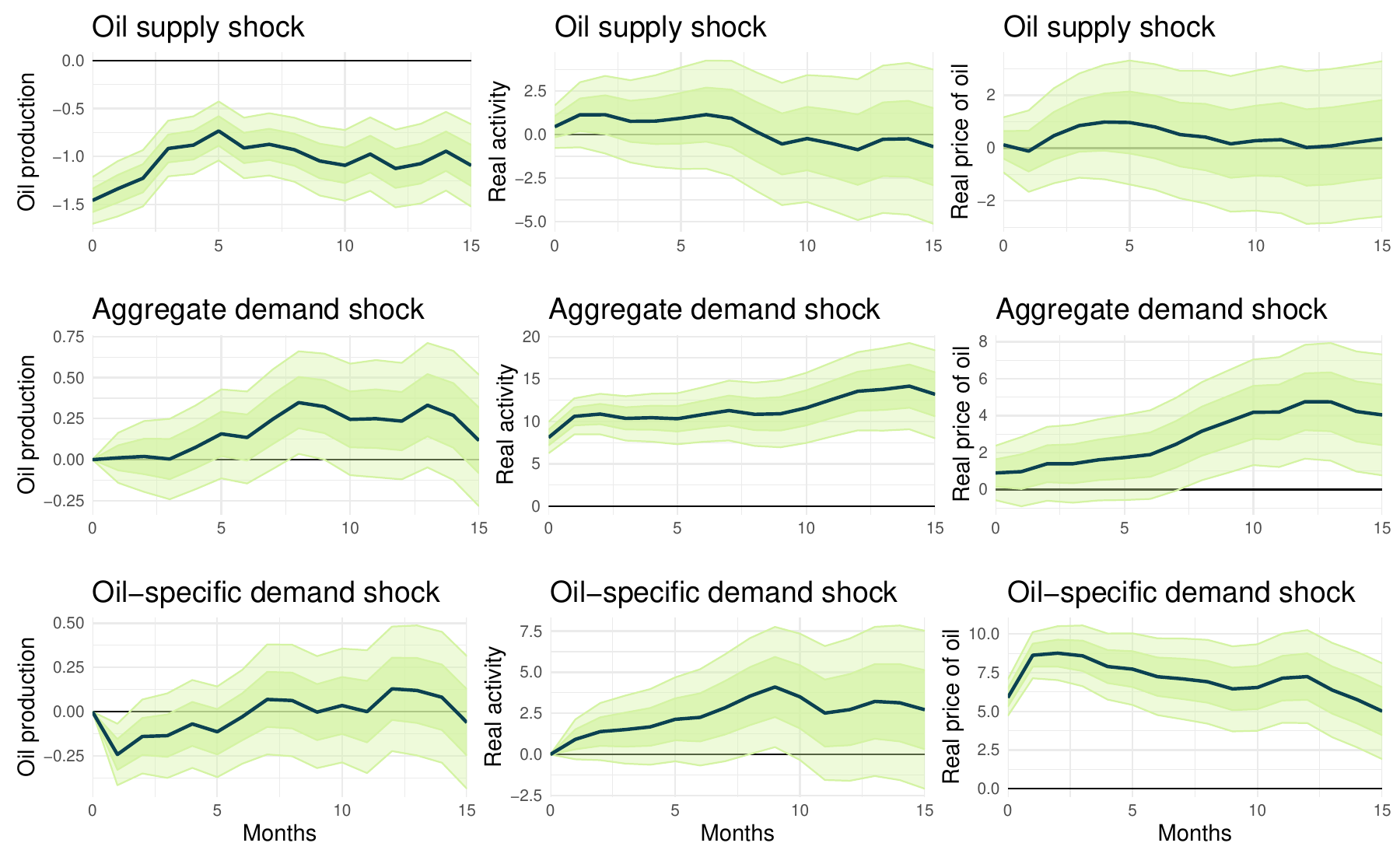}}
\caption[]{\label{fig:irf-update} Responses of one-standard-deviation structural shocks (based on updated data)}
\begin{figurenotes}[Note]
  Solid lines show the point estimates and the shaded regions show one- and two-standard-error bands.
  Estimates are based on the structural VAR model in \eqref{eq:svar},  
  using the updated data (\dateUpdateStart--\dateUpdateEnd).
  Confidence intervals were constructed using a recursive-design wild bootstrap.
\end{figurenotes}
\end{figure}

Figure \ref{fig:irf-update} uses our code base and our data over the period that overlaps the original sample period.
There are three differences to keep in mind when comparing figure \ref{fig:irf-replication} to figure \ref{fig:irf-update}.
First, for the analysis in figure \ref{fig:irf-update},
all the data were retrieved by our code from the sources listed in the text.
Second, the updated data on production start in \dateUpdateStart.
Third,
the data on global real economic activity have been corrected \citep{kilian_2019}.
Notwithstanding these differences,
the patterns are strikingly similar.

An unexpected disruption to oil supply causes an immediate decline in oil production.
The decline is partially reversed within the first 5 months, and
``this pattern is consistent with the view that oil supply contractions 
in one region 
tend to trigger production increases elsewhere in the world''
\citep[][1061--1062]{kilian_2009}.
The fall in production causes a rise in the price of oil, but the 
rise is not overwhelmingly statistically significant.
In figure \ref{fig:irf-update},
the disruption has little effect on real economic activity,
which differs slightly from \citeauthor{kilian_2009}'s \citeyearpar{kilian_2009} observation that a supply disruption
``causes a small temporary reduction of real economic activity that is partially 
statistically significant'' (1,062).

Turning to the second row of figure \ref{fig:irf-update},
the effect of an unanticipated rise in aggregate demand 
causes oil production to rise about 6 months after the shock.
The expansion lasts until 15 months after the shock.
The unanticipated rise in aggregate demand causes a significant rise in global real economic activity.
The effect is persistent.
Expansions in real economic activity cause a significant rise in the real price of oil.
The real price of oil rises on impact (although not statistically significantly) and
the rise continues for a year.

The third row of figure \ref{fig:irf-update} shows 
the effects of an unanticipated rise in oil-specific demand.
An unanticipated rise causes oil production to briefly fall, but 
over the 15-month horizon,
oil-specific-demand increases do not cause oil production to change much.
Real economic activity in response to the unanticipated rise in oil-specific demand expands---albeit modestly, if at all.
What is most clear, though, is that 
the unanticipated rise in oil-specific demand causes the
real price of oil to jump upward immediately and continue to rise for a few months,
which is consistent with models of precautionary demand.
Within this class of models, this overshooting feature is discussed by \citet{kilian_2009} and \citet[][figure 3b, 561]{alquist_kilian_2010}.
Finally, the rise in the real price of oil is highly statistically significant.

Our results are entirely consistent with the results in \citet{kilian_2009}.

\subsection{The cumulative effect of oil demand and oil supply shocks on the real price of oil}
\label{sec:hdecomp}

How much each structural shock explains 
of the historically observed price of crude oil 
can be assessed through a historical decomposition of the covariance-stationary VAR model.
The historical decomposition decomposes $y_t$ into structural innovations through its moving-average representation.
It
attributes the cumulative effect of the
structural innovations in oil production, real economic activity, and oil-specific demand
to the real price of oil at each point in time \citep{kilian_lutkepohl_2017}.

\begin{figure}[htbp]
\centerline{\includegraphics[width=\textwidth]{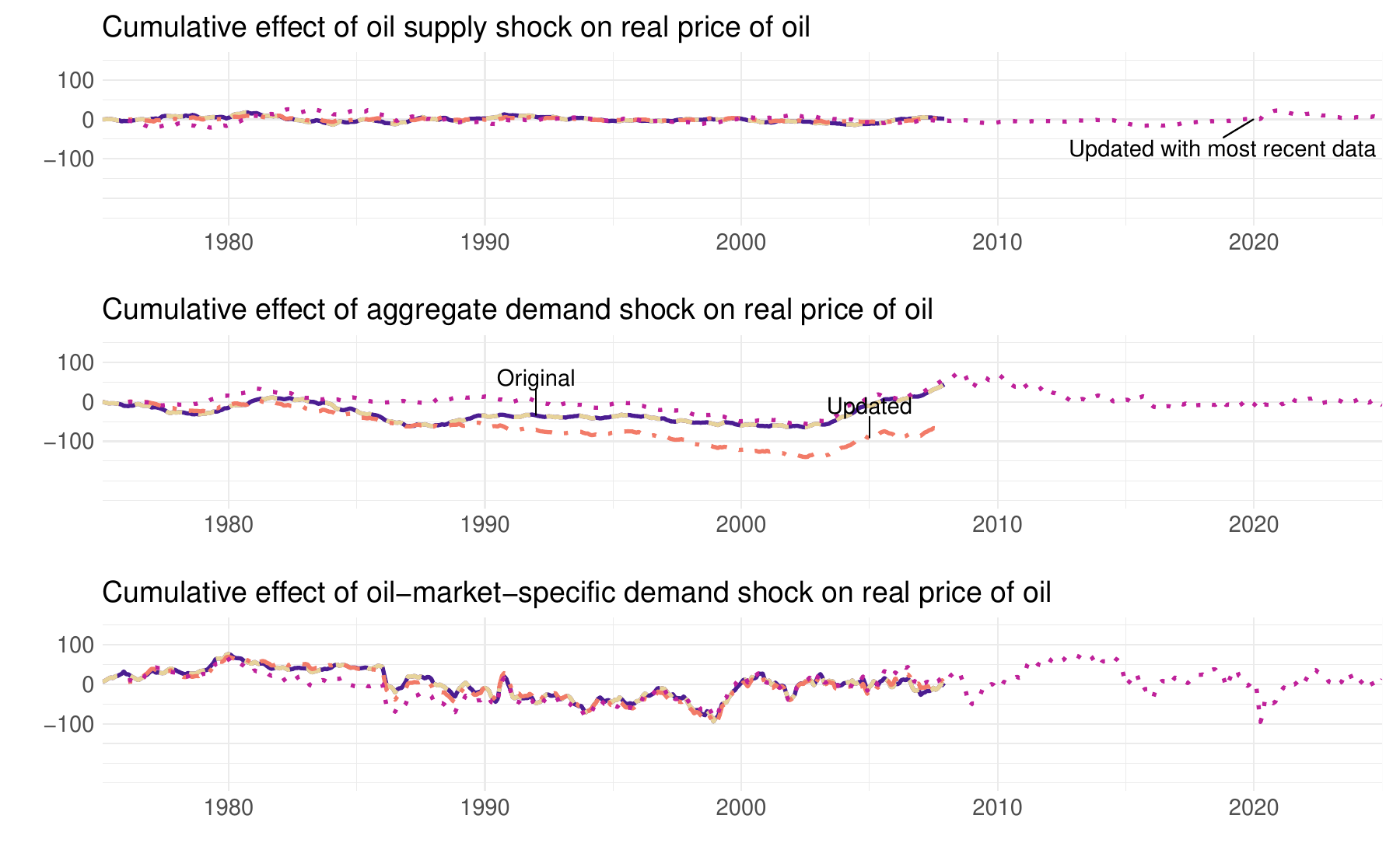}}
\caption[]{\label{fig:hist-decomp} Historical decomposition of the real price of oil}
\begin{figurenotes}[Note]
  \textit{Top}, how much oil-supply shocks explain of the historically observed fluctuations in the real price of crude oil;
  \textit{middle}, how much aggregate-demand shocks explain of the historically observed fluctuations in the real price of crude oil;
  \textit{bottom}, how much oil-specific-demand shocks explain of the historically observed fluctuations in the real price of crude oil.
  Estimates are based on the structural VAR model in \eqref{eq:svar}.
  The solid, purple line is generated by the original code and data.
  The dashed, beige line, which lies on top of the purple line, is generated by the replication code and the original data. 
  The dot--dashed, orange line is generated by the replication code using data from \dateUpdateStart{} to \dateUpdateEnd.
  The dotted, pink line is generated by the replication code using data through \dateFullEnd.
\end{figurenotes}
\end{figure}

Figure \ref{fig:hist-decomp} shows
the cumulative contribution of
oil-supply, aggregate-demand, and oil-specific-demand shocks 
to the real price of oil in separate panels.
Each panel shows 4 lines.
The top panel shows the cumulative effect of oil-supply shocks.
The solid, purple line is generated from the original code and original data.
The dashed, beige line is generated by using the replication code together with the original data.
The two series are numerically equal,
which can be seen by the striped pattern,
the effect of the two series overlapping.
The dot--dashed, orange line is generated by using our replication code and the updated data
through 
\dateUpdateEnd.
The dotted, pink line is generated by using the replication code and updated data through
\dateFullEnd.

The top panel of figure \ref{fig:hist-decomp} shows that
``oil supply shocks historically have made comparatively small contributions to the price of oil'' \citep[][1062]{kilian_2009}.

The middle panel of figure \ref{fig:hist-decomp} shows
that aggregate-demand shocks cause prolonged and persistent changes in the price of oil.
The bottom panel of \ref{fig:hist-decomp} shows
that shocks to oil-specific demand are responsible for
the fluctuations in the price of oil.
As \citet{kilian_2009} points out,
this statistical understanding of crude-oil prices 
is inconsistent with narrative hypotheses about supply disruptions in the Middle East.
There may be supply disruptions, but
the disruptions affect firms that purchase oil, fearing further supply disruptions.
These precautionary-demand motives cause the price of oil to fluctuate.

Figure \ref{fig:hist-decomp} is directly comparable to \citeauthor{kilian_2009}'s \citeyearpar{kilian_2009} figure 4.
The conclusions drawn from both figures agree. 
Therefore, going forward
results will be shown for estimates produced using the full sample.

\subsection{Demand and supply shocks in the global oil market affect the US economy in different ways}
\label{irf2}

How does the global market for oil affect economic activity in the United States?
Having identified the structural innovations using the model in \ref{eq:svar},
it is possible to relate the innovations to US statistics
like the inflation rate measured by the
Consumer Price Index (CPI) and the growth rate of real gross domestic product (GDP).

Because data on real GDP  are available at a quarterly frequency,
the structural shocks are averaged by quarter.
Let $\hat{\zeta}_{jt}$ be the $j$th structural shock in period $t$,
where $t$ now indexes quarter and $j \in \left\{ \text{oil supply, aggregate demand, oil-specific demand} \right\}$.
Put differently, $\hat{\zeta}_{jt}$ is the quarterly average of the monthly values $\hat{\varepsilon}_{j\tau}$,
where $\tau$ is a month within quarter $t$.
\citet{kilian_2009} formalizes the relationship between the structural innovations and
US economic activity by running regressions of the form
\begin{equation}
  z_t = \omega_j + \sum_{i=0}^{12} \phi_{ji} \hat{\zeta}_{jt-i} + u_{jt}, 
  \label{eq:reg2}
\end{equation}
where $z_t$ is either 
the quarterly growth rate of real GDP or
the quarterly inflation rate and
$u_{jt}$ is a potentially serially correlated error.
As \citet{kilian_2009} notes,
the term $\phi_{jh}$  in the regression model is the 
impulse response at horizon $h$ of $z$ to structural innovation $j$.

The possibility for serial correlation in \eqref{eq:reg2} is accounted for
by constructing confidence intervals using a block-bootstrap procedure.
The results presented here use a block length of 6 quarters,
whereas \citet{kilian_2009} uses 4 quarters.

\begin{figure}[htbp]
\centerline{\includegraphics[width=\textwidth]{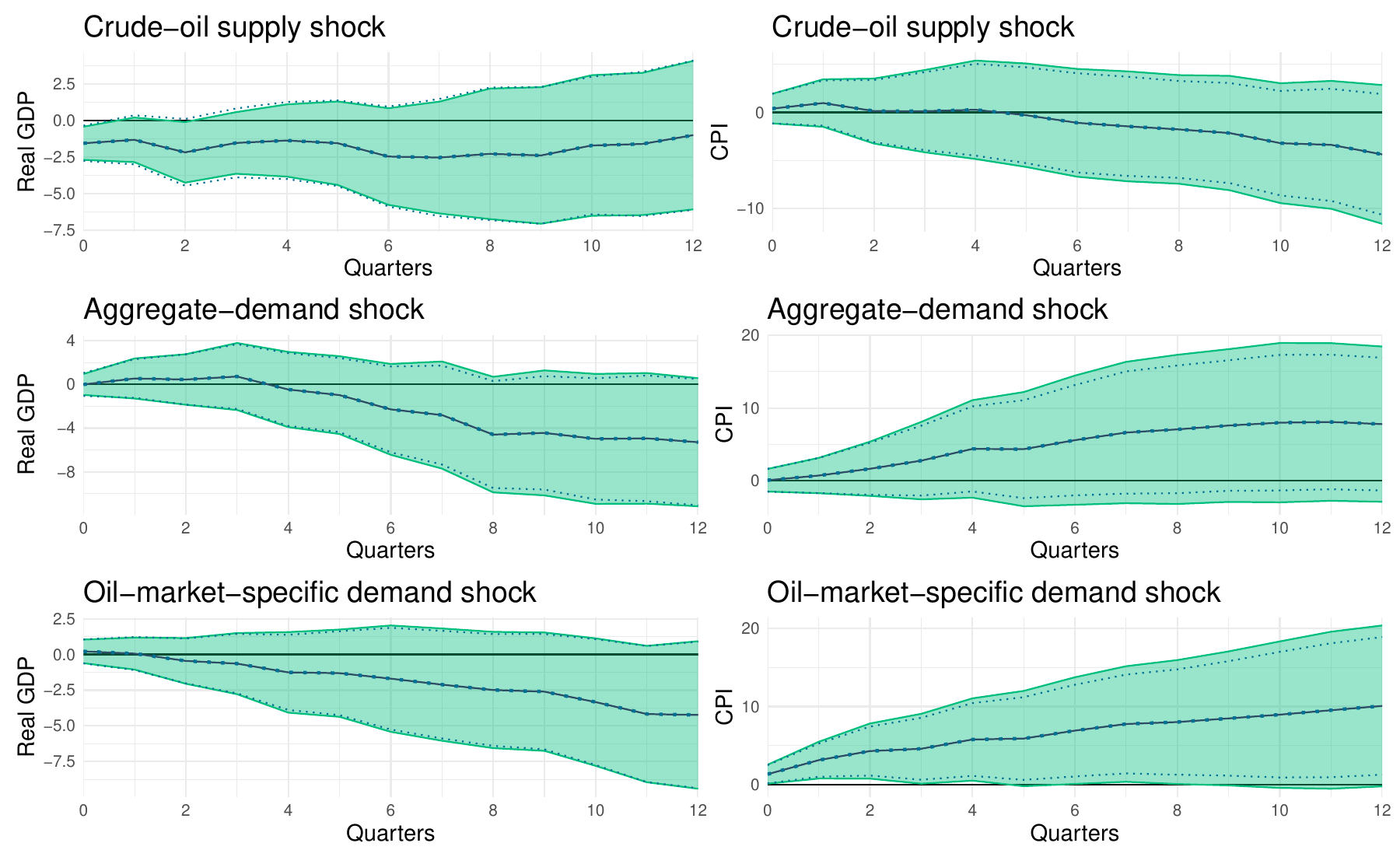}}
\caption[]{\label{fig:irf2} Responses of US real GDP and the CPI to identified structural shocks (based on original data)}
\begin{figurenotes}[Note]
  \textit{Clockwise from top right},
  cumulative response of the price level to an unanticipated disruption in crude-oil supply;
  cumulative response of the price level to an unanticipated increase in aggregate demand;
  cumulative response of the price level to an unanticipated increase in oil-specific demand;
  cumulative response of real GDP to an unanticipated increase in oil-specific demand;
  cumulative response of real GDP to an unanticipated increase in aggregate demand;
  cumulative response of real GDP to an unanticipated disruption in crude-oil supply.
  Solid lines are based on the point estimates from the model in \eqref{eq:reg2}.
  Shaded areas show two-standard-error bands constructed using a block bootstrap procedure.
  Dotted lines show the point estimates and two-standard error bands produced by \citeauthor{kilian_2009}'s \citeyearpar{kilian_2009} code.
  All statistics in the figure are based on the original data \citep{kilian_2009data}.
\end{figurenotes}
\end{figure}

Figure \ref{fig:irf2} shows how the level of US real GDP and 
the level of the CPI respond to the structural shocks.
The solid, gray lines show impulse responses 
estimated using the original data and the replication code.
The associated two-standard-error bands are shaded in green.
The impulse responses estimated using \citeauthor{kilian_2009}'s \citeyearpar{kilian_2009}
code agree exactly,
which produces a striped effect.
The dotted, blue lines show the upper and lower ends of the confidence intervals
using \citeauthor{kilian_2009}'s \citeyearpar{kilian_2009} code.
These estimates produce slightly narrower confidence intervals,
likely due to the fact that the block length used to produce the shaded region
is two quarters longer.
Nevertheless, the difference is minor.
Figure \ref{fig:irf2} is directly comparable to \citeauthor{kilian_2009}'s \citeyearpar{kilian_2009} figure 5.

\begin{figure}[htbp]
\centerline{\includegraphics[width=\textwidth]{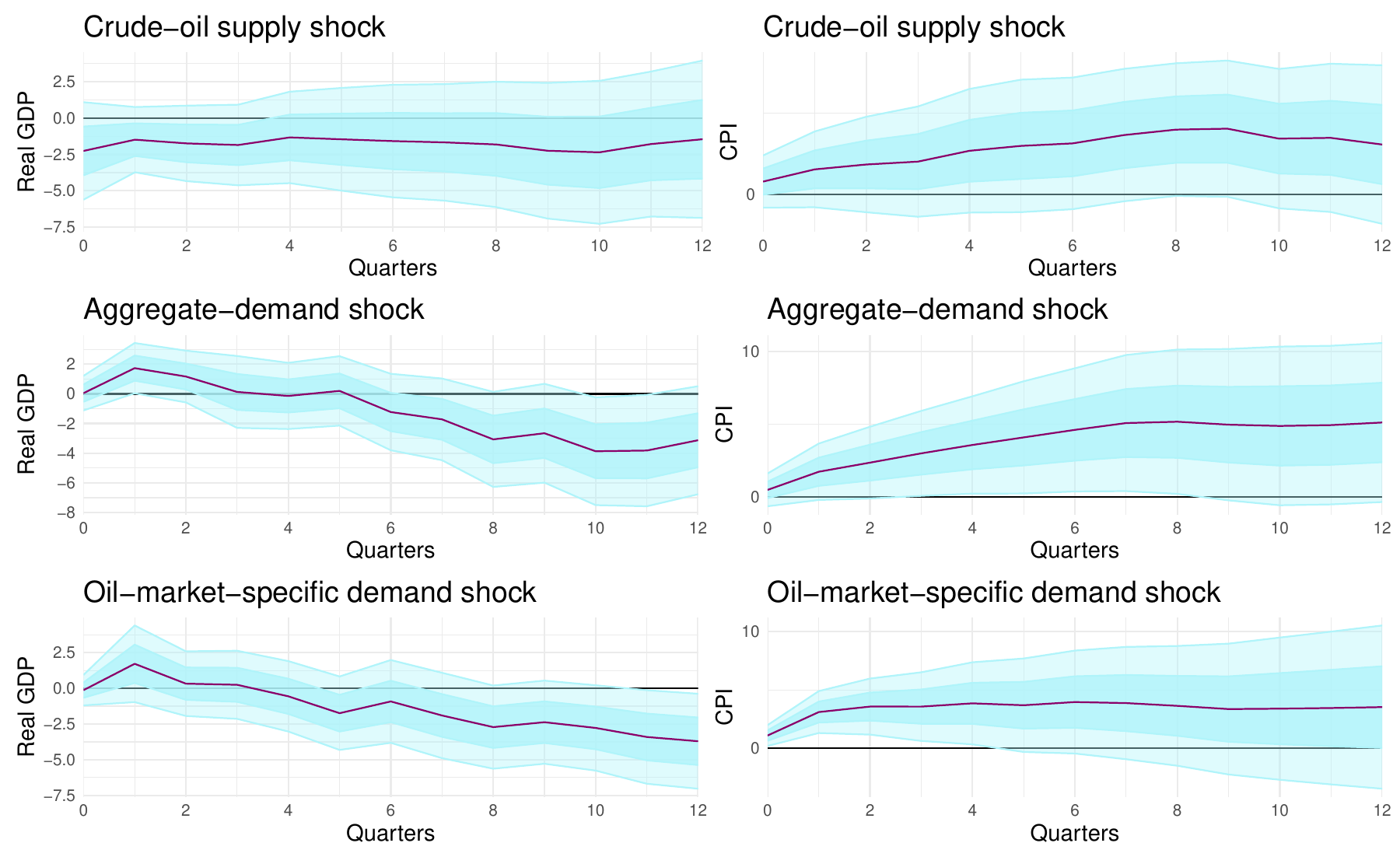}}
\caption[]{\label{fig:irf2-update} Responses of US real GDP and the CPI to identified structural shocks}
\begin{figurenotes}[Note]
  \textit{Clockwise from top right},
  cumulative response of the price level to an unanticipated disruption in crude-oil supply;
  cumulative response of the price level to an unanticipated increase in aggregate demand;
  cumulative response of the price level to an unanticipated increase in oil-specific demand;
  cumulative response of real GDP to an unanticipated increase in oil-specific demand;
  cumulative response of real GDP to an unanticipated increase in aggregate demand;
  cumulative response of real GDP to an unanticipated disruption in crude-oil supply.
  Solid lines are based on the point estimates from the model in \eqref{eq:reg2}.
  Shaded areas show one- and two-standard-error bands constructed using a block bootstrap procedure.
  The estimated structural shocks used to produce the responses are based on data from \dateFullStart{} to \dateFullEnd.
\end{figurenotes}
\end{figure}

Figure \ref{fig:irf2-update} updates figure \ref{fig:irf2}.
To do this,
data from
\dateFullStart{} to 
\dateFullEnd{}
are first used to estimate the structural VAR model.
Then the associated monthly series of structural shocks are averaged by quarter
to produce the quarterly averages used in the second-stage regressions in \eqref{eq:reg2}.

Figure \ref{fig:irf2-update} shows that similar patterns emerge
when
the series for real economic activity is corrected and
the latest data are used to construct estimates.
An unanticipated oil-supply disruption lowers real GDP on impact.
The effect is statistically significant when using the one-standard-error bands
for up to nearly three years.
The unanticipated oil-supply disruption causes the CPI to increase for nearly 9 quarters.
The effect is statistically significant using the one-standard-error bands.
Only after 9 quarters does the price level start to decline.

Turning to the second row of figure \ref{fig:irf2-update},
an unanticipated expansion in aggregate demand
causes real GDP to increase on impact,
although the effect is marginally statistically significant.
After 5 quarters, real GDP falls below its initial level.
The same unanticipated expansion causes the price level to rise 
for the first 7 quarters after the shock.
The response is statistically significant using the one-standard-error bands.
This pattern is consistent with
the increase in global aggregate demand causing an initial increase in US growth, but
the growth is eventually hampered by the higher price of global commodities, including the price of oil,
causing real GDP to fall.
\citet{kilian_2009} provides further discussion.

The third row of figure \ref{fig:irf2-update}
shows that an unanticipated increase in oil-specific demand 
causes real GDP and the price level to rise.
The increase in the price level is highly statistically significant.
Eventually the higher prices cause real GDP to fall.
Again, \citet{kilian_2009} provides further discussion.

In summary,
we are able to replicate \citeauthor{kilian_2009}'s \citeyearpar{kilian_2009} substantive results.
Patterns uncovered by 
the historical decomposition and
the second-stage impulse response functions
hold using data updated through 
\dateFullEnd.

\section{Updated responses of oil production, real economic activity, and the real price of oil to demand and supply shocks}
\label{sec:update}

Section \ref{sec:hdecomp} provides evidence
that conclusions about
the determinants of the real price of crude oil apply
through \dateFullEnd;
namely,
disruptions to oil supply contribute little to fluctuations in price,
whereas
oil-specific demand shocks are responsible for the ups and downs and
aggregate-demand shocks cause ``long swings'' in price
\citep[][1062]{kilian_2009}.
Section \ref{irf2} provides evidence
that the US economy responds similarly to changes in the oil market over the later period.

In this section we use updated data to estimate the impulse response functions.
We follow \citet{kilian_lutkepohl_2017} and construct confidence intervals using the moving-block-bootstrap procedure introduced by \citet{bruggemann_jentsch_trenkler_2016}. 
\citet[][pp 415--416]{kilian_lutkepohl_2017} use this procedure to assess \citeauthor{kilian_2009}'s \citeyearpar{kilian_2009} conclusions. 
(We are doing nothing new---we are using the same inference procedures, only inference is being carried out over a different sample period.) 
\citeauthor{bruggemann_jentsch_trenkler_2016}'s \citeyearpar{bruggemann_jentsch_trenkler_2016} recursive-design residual-block-bootstrap procedure is robust to conditional heteroskedasticity. 
In addition,
the procedure is designed to replicate the asymptotic distribution
of both the slope parameters and the variance--covariance matrix.
In effect, the wild bootstrap ``produces confidence intervals that
are too narrow and have coverage accuracy that is too low asymptotically''
\citep[][346--347]{kilian_lutkepohl_2017}.
An excellent discussion of the issue is provided by \citet{kilian_lutkepohl_2017}.
To implement the moving-block bootstrap procedure,
a block length of 36 months is chosen.

\begin{figure}[htbp]
\centerline{\includegraphics[width=\textwidth]{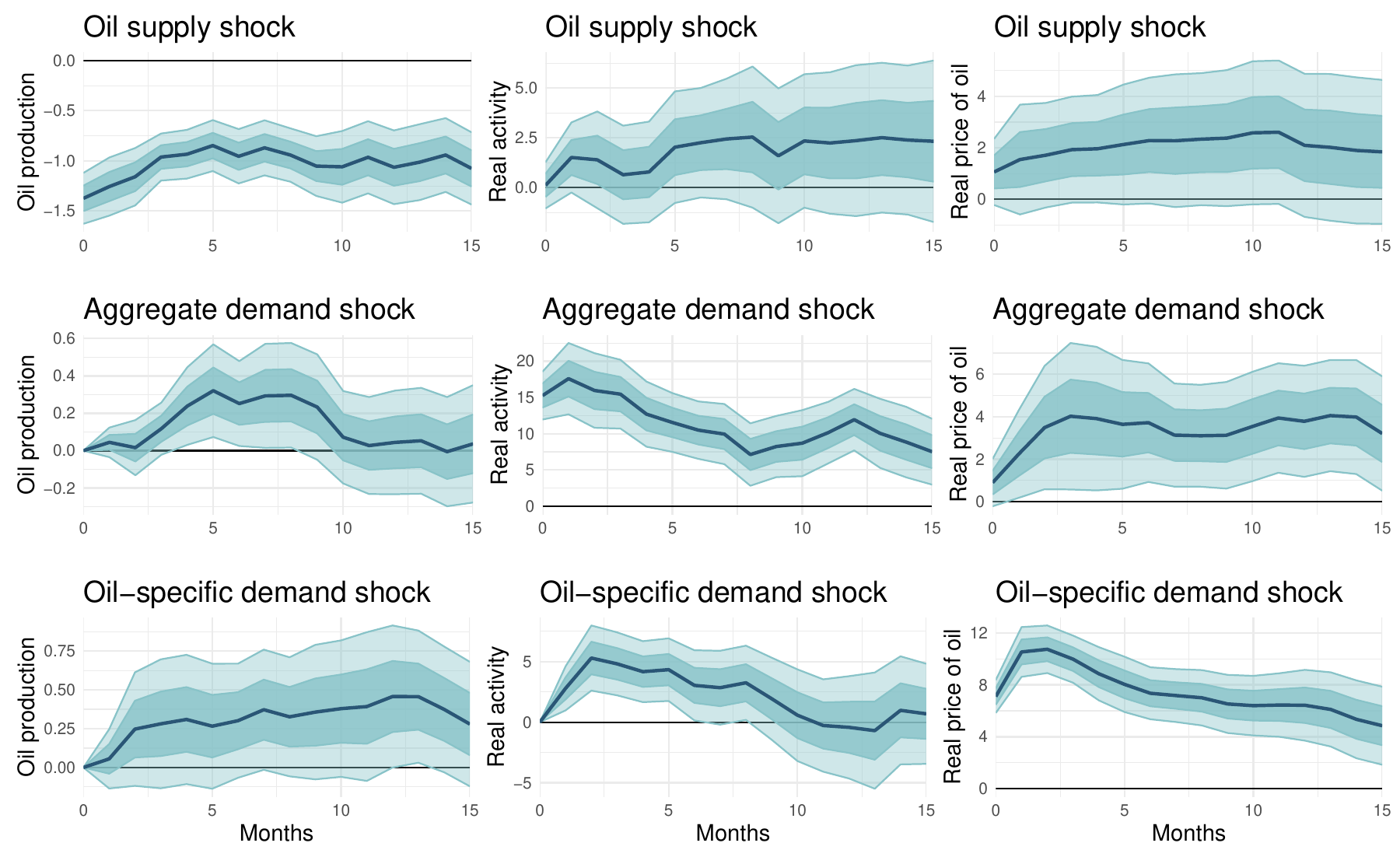}}
\caption[]{\label{fig:irf-full} Responses to one-standard-deviation structural shocks}
\begin{figurenotes}[Note]
  Solid lines show the point estimates and the shaded regions show one- and two-standard-error bands.
  Estimates are based on the structural VAR model in \eqref{eq:svar},  
  using the updated data (\dateFullStart--\dateFullEnd).
  Confidence intervals were constructed using a residual block bootstrap procedure \citep{bruggemann_jentsch_trenkler_2016}.  
\end{figurenotes}
\end{figure}

Figure \ref{fig:irf-full} shows the responses of
production, real economic activity, and the real price of oil to one-standard-deviation structural innovations.
When comparing figure \ref{fig:irf-full} to figure \ref{fig:irf-update},
there are three differences to keep in mind.
First, the structural VAR model in \eqref{eq:svar} is estimated using data retrieved by our code from the sources listed in the text.
Second, 
the data cover the period \dateFullStart{} to \dateFullEnd.
Third, the confidence intervals are constructed using the procedure introduced by \citet{bruggemann_jentsch_trenkler_2016}.
Nevertheless,
results are remarkably similar.
An unexpected disruption to oil supply 
causes an immediate decline in oil production and
the decline persists.
The unexpected fall in oil production initially has no effect on bulk-dry-cargo shipping rates relative to trend, 
but this measure of real economic activity picks up after 4 months.
The rise could reflect higher fuel costs or expected higher fuel costs
(relative to the US consumer price index, which is used to deflate the series).
For example, the next panel shows that
the unexpected disruption to supply causes an even more pronounced rise in the price of oil.
The rise in the price of oil is statistically significant at all horizons
based on the one-standard-error bands.

Turning to the second row of figure \ref{fig:irf-full},
over the entire sample,
oil production seems to respond more immediately to an unanticipated increase in aggregate demand,
although a months-long delay remains.
The unanticipated rise in aggregate demand causes a large, significant jump in global real economic activity,
but the effect is less persistent compared to the effect estimated over only the earlier period.
And, compared to only the earlier period,
there is a more pronounced rise in the real price of oil in response to an 
unanticipated increase in aggregate demand.
Nevertheless,
the patterns are overwhelmingly consistent with \citeauthor{kilian_2009}'s \citeyearpar{kilian_2009} substantive conclusions.

The bottom row of figure \ref{fig:irf-full}
shows that an unanticipated rise in oil-specific demand
causes an immediate rise in the real price of oil.
Again there is evidence of a precautionary-demand motive.
Like over the earlier period,
the unanticipated rise in oil-specific demand 
is also associated with an increase in real economic activity and a rise in oil production.
Perhaps, unlike in the earlier period, 
oil production responds more to surprises in oil-specific demand,
although there is a delay of two months and
the effect is significant using the one-standard-error bands only.

In summary,
the substantive conclusions of \citet{kilian_2009} hold
when 
all available data are used and
confidence sets are constructed using a method
robust to potential conditional heteroskedasticity.

\section{Application to the clean-energy transition}
\label{sec:app}

Much previous interest has centered around the
role that oil-supply and -demand shocks have on macroeconomic outcomes
\citep[for example,][looked at macroeconomic outcomes in response to oil-demand and oil-supply shocks]{kilian_2008,kilian_2009}.
Another compelling reason to study these types of shocks is 
understanding how local economies depend on oil.

Certain local economies may be especially hard hit as the US transitions away from fossil fuels 
to sources of clean energy.
Previous structural transitions---such as the transition away from coal and
the transition to work in the United States being replaced by machines and jobs elsewhere---are causes for concern.
Workers displaced from jobs that are no longer performed 
are especially hard-hit.
And concern extends to the communities in which these workers live \citep{hanson_2022}.\footnote{The forces
  that restructure the composition of economic activity and
  the consequences of these forces are not fully understood.
  Job loss in the coal industry is discussed by \citet{weber_2020}.
  Between 2011 and 2016, the number of people working in mines declined by 45 percent.
  Each lost coal-mining job is associated with a reduction in county-level income by \$100,000 annually.
  Retirement of coal-fired power plants in the United States is discussed by
  \citet{davis_holladay_sims_2022} and
  the negative effects on workers of shutting down coal-fired power plants are discussed by
  \citet{hanson_2022,colmer_etal_2024}.
  Work being automated is discussed by \citet{acemoglu_autor_2011,acemoglu_restrepo_2020,acemoglu_restrepo_2022}.
  World trade and China's participation in world trade is discussed by \citet{autor_dorn_hanson_2021,autor_dorn_hanson_2016}.
  The change in composition associated with the hollowing out of middle-pay jobs and when this reallocation takes place over the business cycle is discussed by \citet{autor_dorn_2013,foote_ryan_2014,howes_2022}.
  \citet{herrendorf_rogerson_valentinyi_2014} provide a framework 
  for structural change within the tradition of economic growth \citep[see also][]{ngai_pissarides_2007}.
  Within this framework,
  \citet{rogerson_2015} provides a way to think about evaluating environmental regulation.}

In California, 
policymakers have adopted the goal of carbon neutrality by 2045 or earlier \citepalias{carb_2022}.
Within California, Kern County accounts for 70 percent of all oil produced.
As California transitions away from oil,
major concerns for Kern include lost jobs and lost tax revenue.
Many workers there are primarily engaged in
the exploration, development, and production of oil.
In addition,
the assessed value of oil-field properties fluctuates with factors 
primarily outside the control of Kern policymakers \citep{kern_2024}. 
In particular, fluctuations in oil futures prices determine valuations used to assess property taxes.
Despite efforts by Kern policymakers to smooth the ups and downs, 
shocks from the global oil market may propagate to other sectors.

Linkages between labor-market outcomes in oil-producing counties and the oil market
have been investigated previously.
\citet{michieka_gearhart_2019,michieka_gearhart_2021} investigate
relationships between nominal regional oil prices and
\begin{enumerate*}[(i)]
\item\label{item:2} employment and 
\item nominal wages 
\end{enumerate*}
in oil-producing counties.
The dynamic-panel models they use are based on the idea that
along with the number of time periods getting large,
the number of cross-sectional groups is also relatively large.
A common ``long-run relationship'' between regional oil prices and other variables
is posited to exist across counties.
Importantly,
they focused on the causal short- and long-run effects of oil prices
(consistent with the econometric specification here that assumes oil prices are exogenous
with respect to outcomes in Kern).
In contrast, we are interested in the impulse response functions,
which trace out the responses of labor-market outcomes in Kern
to one-time structural innovations in the global oil market.
We are also interested in the labor-market dynamics associated with a disruption to supply and
a shift in demand, including 
whether the shift represents a structural innovation to 
global real economic activity or precautionary demand.\footnote{\citet{michieka_gearhart_2015}
  posit a cointegrating relationship between the nominal price of oil and employment in Kern County and
  estimate a vector error correction model.
  The rationale for imposing the cointegrating relationship is statistical rather than economic,
  which is one reason to examine the sensitivity to this modeling assumption.
  \citet{michieka_gearhart_2015} do not investigate impulse responses,
  which is our interest.}

It is nearly impossible to definitively state how much of Kern's economy depends on oil.
Yet, one way to understand the linkages between Kern and the global market for oil 
is by assessing how Kern's labor market 
responds to structural innovations identified by the statistical model in \eqref{eq:svar},
using the second-stage methodology proposed by \citet{kilian_2009}.
Specifically,
we regress the seasonally adjusted
labor-market statistics in Kern
on lags of the estimated structural shocks:
\begin{equation}
\label{eq:kern}
k_{t} = \kappa + \sum_{i=0}^{12} \psi_{ji} \hat{\varepsilon}_{jt-i} + e_{jt},
\end{equation}
where
$k_{t}$ is either the unemployment rate or the monthly growth rate of employment,
$\hat{\varepsilon}_{jt}$
refers to the estimated residual in \eqref{eq:svar} for the $j$th structural shock in month $t$, and
$e_{jt}$ is a potentially serially correlated error.
The coefficient $\psi_{jh}$ is
the impulse response at horizon $h$ of series $k$ to structural innovation $j$.
Inference is based on a block-bootstrap procedure,
where the block length is chosen to be half a year.

Data on the level of employment and the unemployment rate in Kern County, California, are available 
from the Local Area Unemployment Statistics program of the Bureau of Labor Statistics
starting in \dateKernStart.
The data extend through \dateKernEnd.
Both series are retrieved from FRED and
seasonally adjusted using the US Census Bureau's X-13ARIMA-SETS procedure.

\begin{figure}[htbp]
\centerline{\includegraphics[width=\textwidth]{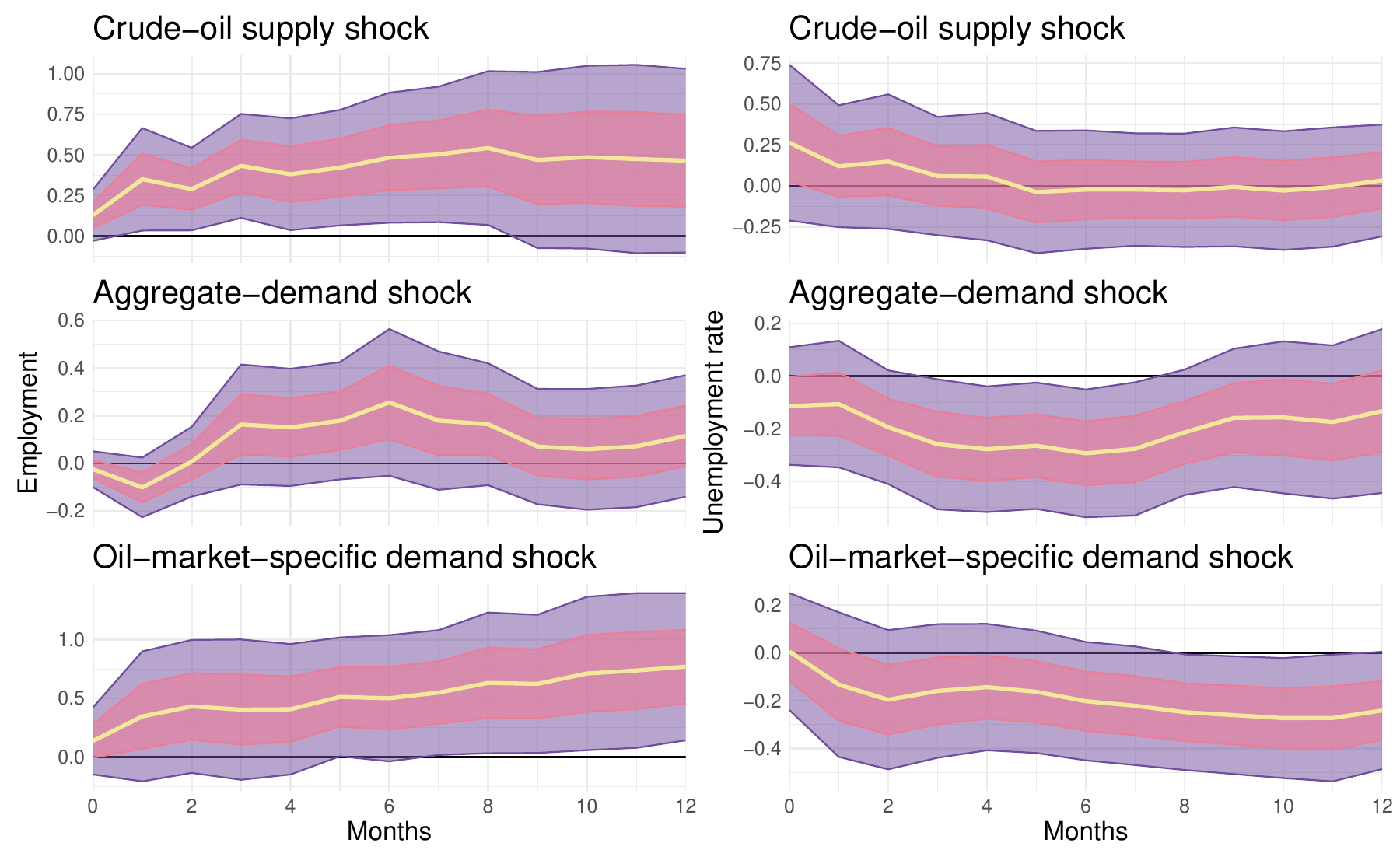}}  
\caption[]{\label{fig:irf2-kern} Responses of employment and the unemployment rate in
  Kern County, California, to identified structural shocks}
\begin{figurenotes}[Note]
  \textit{Clockwise from top right},
  response of the unemployment rate to an unanticipated disruption in oil supply;
  response of the unemployment rate to an unanticipated increase in aggregate demand;
  response of the unemployment rate to an unanticipated increase in oil-specific demand;
  cumulative response of employment to an unanticipated increase in oil-specific demand;
  cumulative response of employment to an unanticipated increase in aggregate demand;
  cumulative response of employment to an unanticipated disruption in oil supply.
  Solid lines are based on the point estimates from the model in \eqref{eq:kern}.
  Shaded areas show one- and two-standard-error bands constructed using a block bootstrap procedure.
  The estimated structural shocks used to produce the responses are based on data from
  \dateFullWIPStart{} to \dateFullWIPEnd.
  The regressions that use Kern's unemployment rate begin in \dateKernStart{} and extend through \dateKernEnd.
  The regressions that use the change in employment begin one month later. 
\end{figurenotes}
\end{figure}

Figure \ref{fig:irf2-kern} shows the estimated coefficients.
The solid, beige lines show the impulse-response coefficients.
The shaded areas show the one- and two-standard-error bands.
Because employment is entered into the statistical model as
the change in log employment,
we sum the estimated coefficients from \eqref{eq:kern} 
to estimate the cumulative response of employment.

The top, left panel shows that,
upon impact,
an unanticipated disruption in oil production causes employment to rise.
This evidence is consistent with the idea that 
``oil supply contractions in one region tend to trigger production increases elsewhere in the world''
\citep[][1060--1062]{kilian_2009}.
The evidence is also consistent with the notion that an increase in the price of oil
represents a windfall boost to income from the perspective of some in Kern's economy.
The windfall may propagate throughout the local economy.
Employment rises and, turning to the top, right panel,
expanded opportunities increase participation in the labor market, as the unemployment rate rises.
The effect on the unemployment rate, however, is
not statistically significant based on the one- and two-standard-error bands.

In contrast, 
an unanticipated one-standard-deviation increase in aggregate demand
causes the unemployment rate to fall on impact.
The effect is statistically significant after 2 months based on the two-standard-error bands.
The effect on employment, however, is less pronounced.
On impact the effect cannot be distinguished from zero.
And while employment expands after two months,
the effect is statistically significant based on one-standard-error bands only.
These conclusions can be seen in the middle panel of figure \ref{fig:irf2-kern}.

The bottom panel shows that
an unanticipated increase in oil-specific demand expands employment opportunities in Kern.
On impact, employment rises.
And continues to rise until by 7 months the effect is
statistically different from zero based on the two-standard-error bands.
Employment is higher a year after the initial shock.
The right panel shows that the unemployment rate falls on impact.
The effect is statistically different from zero after around 9 months.
These effects are consistent with the notion that
oil production in Kern may respond to oil-specific demand and
local revenue from oil production will rise.
Linkages between oil extraction and the regional economy propagate to
expanded opportunities throughout Kern's labor market.

The effects suggest that Kern depends on oil in a meaningful way.
As the US economy transitions away from dependence on fossil fuels toward green energy,
Kern's economy may be drastically reshaped.
What had previously been
a major component of regional economic activity may no longer provide opportunities.
Policymakers may need to consider
expanding options in Kern through, for example, place-based policy.
Further research on this topic will 
be aided by having the option to work within the R ecosystem.

Figure \ref{fig:irf2-kern} makes clear, though,
that \citeauthor{kilian_2009}'s \citeyearpar{kilian_2009} point about identifying the origin of the shock
matters.
Even though the price of oil may be determined independently of economic activity in Kern County,
the dynamic effect of a negative supply disruption differs from
the dynamic effect of a positive demand innovation.
A negative supply disruption expands labor-market opportunities more rapidly than the
expanded opportunities associated with an oil-specific-demand shock.
The effects of a positive demand innovation are visible in the data after 1 year.

\section{Conclusion}
\label{sec:conclusion}

In this paper we successfully replicated the substantive conclusions made by \citet{kilian_2009}; namely,
not all oil-price shocks are alike.
Understanding the origins of such shocks is necessary for policymakers.
As \citet{kilian_2009} points out,
a positive structural innovation to the global business cycle
will cause national income in the United States to increase---as the 
United States directly benefits from an increase in global economic activity---but
the expansion of US income will be hampered by the rising cost of oil that inevitably follows.
A negative innovation to crude-oil production will cause national income in the US to fall and cause
the CPI to rise.
Both innovations initially cause the price of crude oil to rise, but
the dynamic effects differ meaningfully.
Policymakers will need to understand the origin of the shock in order to respond appropriately.

\citet{kilian_2009} should be commended for his excellent code.
The original replication package reproduces the results even 
when we port the code to R \citep{kilian_2009data}. 
In fact,
the original replication package offers an excellent starting point for further analysis.

We furthered the analysis in several ways:

\begin{itemize}
\item We ported the code to R,
  which will benefit researchers interested in working within the R ecosystem.
  For many, tidying data in R is much easier than tidying data in MATLAB.
\item Our replication and paper are written using the \texttt{targets}
  \href{https://books.ropensci.org/targets/}{package}.
We demonstrate how to use this Make-like took to ensure reproducibility.
The \texttt{targets} package can also be used to exploit parallel computing,
which could be used to speed the construction of bootstrapped confidence intervals.
\item Data from the US Energy Information Administration and the FRED database
  are retrieved by code.
  This makes updating the structural shocks straightforward.
\item We provided an application of \citeauthor{kilian_2009}'s \citeyearpar{kilian_2009} procedure for identifying structural shocks.
  These shocks are used learn about how a local labor market responds to the global oil market.
  Understanding linkages like these may well be more important as society transitions toward sustainable energy sources.
\end{itemize}

\bibliographystyle{../../../bibliography/bostonfed}
\bibliography{../../../bibliography/bibliography-org-ref}
\appendix
\section{Appendix}

\subsection{Data}
\label{sec:data}

\citeauthor{kilian_2009}'s \citeyearpar{kilian_2009} structural VAR model includes three variables:
\begin{enumerate*}[(1)]
\item\label{item:3} the log difference of global crude-oil production,
\item a measure of cyclical variation in global real economic activity, and
\item the log of the real price of oil.
\end{enumerate*}
Data on production and price are available from the 
\href{https://www.eia.gov/}{US Energy Information Administration} (EIA).
The index of global real economic activity is made available by the
Federal Reserve Bank of Dallas through the FRED database.
Details about accessing the data are discussed below.

Data on production are available from the EIA under 
\href{https://www.eia.gov/international/data/world}{Geography}.\footnote{Using
  the \href{https://www.eia.gov/international/data/world}{link} in the text, the data can be found under ``Petroleum and Other Liquids'' and
    ``Crude oil including lease condensate.''}
Data on price are also available from the EIA under the series
\href{https://www.eia.gov/totalenergy/data/browser/index.php?tbl=T09.01}{Refiner Acquisition Cost of Crude Oil, Imported}.\footnote{Further details are provided in Section 9 of the EIA's \textit{Monthly Energy Review}.} 
The EIA website includes links for retrieving the data through a web
application programming interface or API.
Through the API, it is possible
to send structured HTTP requests that return data with JSON format or
\textbf{J}ava\textbf{S}cript \textbf{O}bject \textbf{N}otation format.
The code that gathers our data
uses the EIA's API to retrieve the latest data for production and price.

The nominal price series is converted to constant dollars using the Consumer Price Index for All Urban Consumers.\footnote{U.S.~Bureau of Labor Statistics,
  Consumer Price Index for All Urban Consumers: All Items in U.S. City Average [CPIAUCSL], retrieved from FRED, Federal Reserve Bank of St. Louis;
  \href{https://fred.stlouisfed.org/series/CPIAUCSL}{\nolinkurl{https://fred.stlouisfed.org/series/CPIAUCSL}}.}

The time series used as an indicator of the global business cycle is
the Index of Global Real Economic Activity.
The series is available through FRED.\footnote{Federal Reserve Bank of Dallas, Index of Global Real Economic Activity [IGREA], 
  retrieved from FRED, Federal Reserve Bank of St. Louis;
  \href{https://fred.stlouisfed.org/series/IGREA}{\nolinkurl{https://fred.stlouisfed.org/series/IGREA}}.}
It ``is derived from a panel of dollar-denominated global bulk dry cargo shipping rates and may be viewed as a proxy for the volume of shipping in global industrial commodity markets.''
\citet{kilian_2009} initially used this measure of the global business cycle.
\citet{kilian_2019} reviews the index and
\citet[][55]{kilian_zhou_2018} discuss the merits of the series, concluding
``that the \citet{kilian_2009} index and indices based on real commodity prices have clear advantages in modeling industrial commodity markets compared with alternative indicators of global real economic activity, including proxies for global industrial production.''

\begin{figure}[htbp]
\centerline{\includegraphics[width=\textwidth]{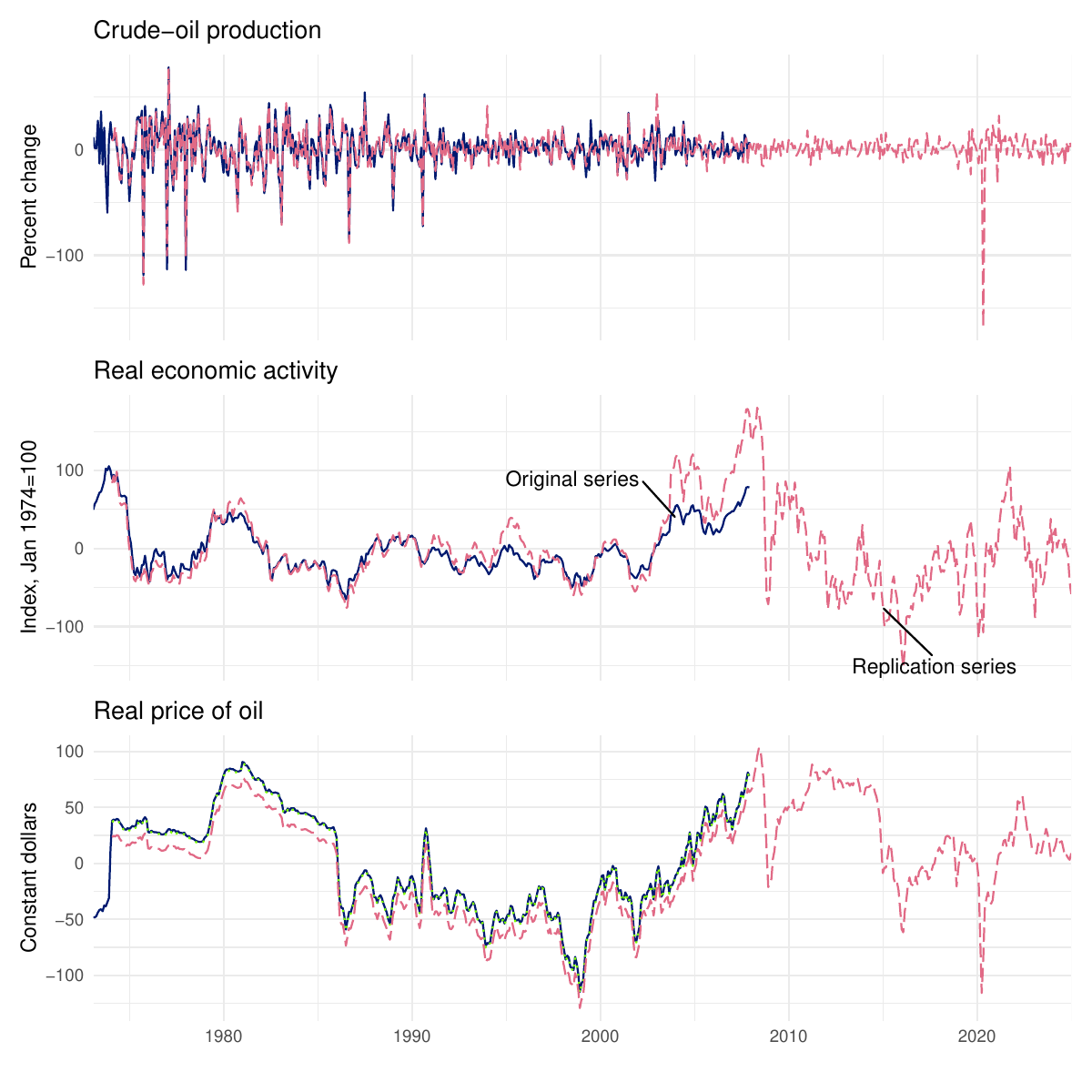}}
\caption[]{\label{fig:var-data} Determinants of the global market for crude oil}
\begin{figurenotes}[Note]
  \textit{Top}, change in global crude-oil production;
  \textit{middle}, global real economic activity measured
  using the index proposed by \citet{kilian_2009} and revised by \citet{kilian_2019}; 
  \textit{bottom}, real price of oil measured by
  the acquisition cost of refiners and deflated using the Consumer Price Index for All Urban Consumers.
  The solid, blue lines show the data originally used by \citet{kilian_2009}.
  The dashed, red lines show the updated data.
  The dotted, green line in the bottom panel shows the updated series for the real price of oil
  demeaned over the period \dateFullStart{} to \dateOrigEnd,
  which overlaps the original sample period after 1974.
  The striped effect indicates the two series coincide.
  All data are monthly.
  The updated data begin in \dateFullStart{} and go through \dateFullEnd.
  The original data begin one year earlier.  
\end{figurenotes}
\begin{figurenotes}[Sources]
  US Energy Information Administration, Federal Reserve Bank of Dallas, and US Bureau of Labor Statistics.
\end{figurenotes}
\end{figure}

Figure \ref{fig:var-data} depicts the three series used in the structural VAR model.
All series are monthly.
The top panel depicts the change in global crude-oil production.
The solid, blue line shows the data used by \citet{kilian_2009}, and
the dashed, red line shows data retrieved using the code for this paper
\citep[see][for the original data and code]{kilian_2009data}.
The original series begin in 
\dateOrigStart{}
and go through
\dateOrigEnd.
The updated series used for the replication begin in
\dateFullStart{}
and go through
\dateFullEnd.
\citet{kilian_2009} extends the series through 1973 using 
a procedure described by \citet{barsky_kilian_2001},
a procedure that is not undertaken here.
The two series largely coincide.
The striped effect indicates that the series overlap. 

The middle panel in figure \ref{fig:var-data} shows the time series of global real economic activity.
Again,
the solid, blue line depicts the original data and
the dashed, red line depicts the updated data.
A few years before 2005,
the two series diverge. 
This discrepancy has to do with a coding error in the original analysis described by \citet{kilian_2019}.

The bottom panel in figure \ref{fig:var-data} shows the time series for the real price of crude oil.
The solid, blue line shows the series used by \citet{kilian_2009}.
The dashed, red line shows data retrieved by our code.
The discrepancy between the series is accounted for by demeaning over different periods.
The dotted, green line shows our updated data used in the replication demeaned over
the period that overlaps with the original data.
The striped effect indicates the two series agree.

As \citet{kilian_2009} demonstrates,
the three time series in figure \ref{fig:var-data} are key 
determinants of the global market for crude oil.
These are the three series used to estimate the structural VAR in the paper.

\subsection{A sensitivity analysis that uses an index of global industrial production}
\label{sec:world-ind-prod}

\citet{kilian_2009} uses a measure of global shipping rates as an indicator of global real economic activity,
which reflects global demand for all industrial commodities \citep[][provides an update]{kilian_2019}.
We focus on \citeauthor{kilian_2009}'s \citeyearpar{kilian_2009} index in the text for two reasons.
First,
it was used in \citeauthor{kilian_2009}'s \citeyearpar{kilian_2009} original specification of the oil market and
replicating that analysis was our goal.
Second,
\citet{kilian_zhou_2018} make a persuasive case that \citeauthor{kilian_2009}'s \citeyearpar{kilian_2009} index is the appropriate index to use in this application:
\citeauthor{kilian_2009}'s \citeyearpar{kilian_2009} index has
``the advantage of being [a leading indicator] with respect to changes in latent global real output and of accounting for the role of expectations about the business cycle''
\citep[][75]{kilian_zhou_2018}.

An alternative indicator of global real economic activity is the OECD+6 index of industrial production.
This is an index of industrial production for
OECD (Organization for Economic Co-operation and Development) countries and, starting in 2006,
six non-member countries (Brazil, China, India, Indonesia, the Russian Federation, and South Africa).
It serves as a proxy for global industrial production.
\citet{kilian_zhou_2018} point out that
it is a coincident indicator (as opposed to a leading indicator) and it fails to account for Chinese demand prior to 2006.
The OECD+6 industrial production index was first explored by \citet{baumeister_kilian_2014} and
has been used by \citet{baumeister_hamilton_2019}.
For further discussion see \citet{kilian_2022}.

We follow \citet{zhou_2020},
who conducts a similar sensitivity analysis for a refinement of \citeauthor{kilian_murphy_2014}'s \citeyearpar{kilian_murphy_2014} model of the global oil market, and 
replace \citeauthor{kilian_2009}'s \citeyearpar{kilian_2009} index with the
first difference of the log of the OECD+6 index.
\setcitestyle{square}
(\citet{zhou_2020} uses the log-linearly detrended OECD+6 index.) 
\setcitestyle{round}

For our analysis,
we use the index published on Christiane Baumeister's website.\footnote{Which is
  accessible at \href{https://sites.google.com/site/cjsbaumeister/datasets}{\nolinkurl{https://sites.google.com/site/cjsbaumeister/datasets}}.}

\subsubsection{Impulse responses based on the structural VAR}

\begin{figure}[htbp]
\centerline{\includegraphics[width=\textwidth]{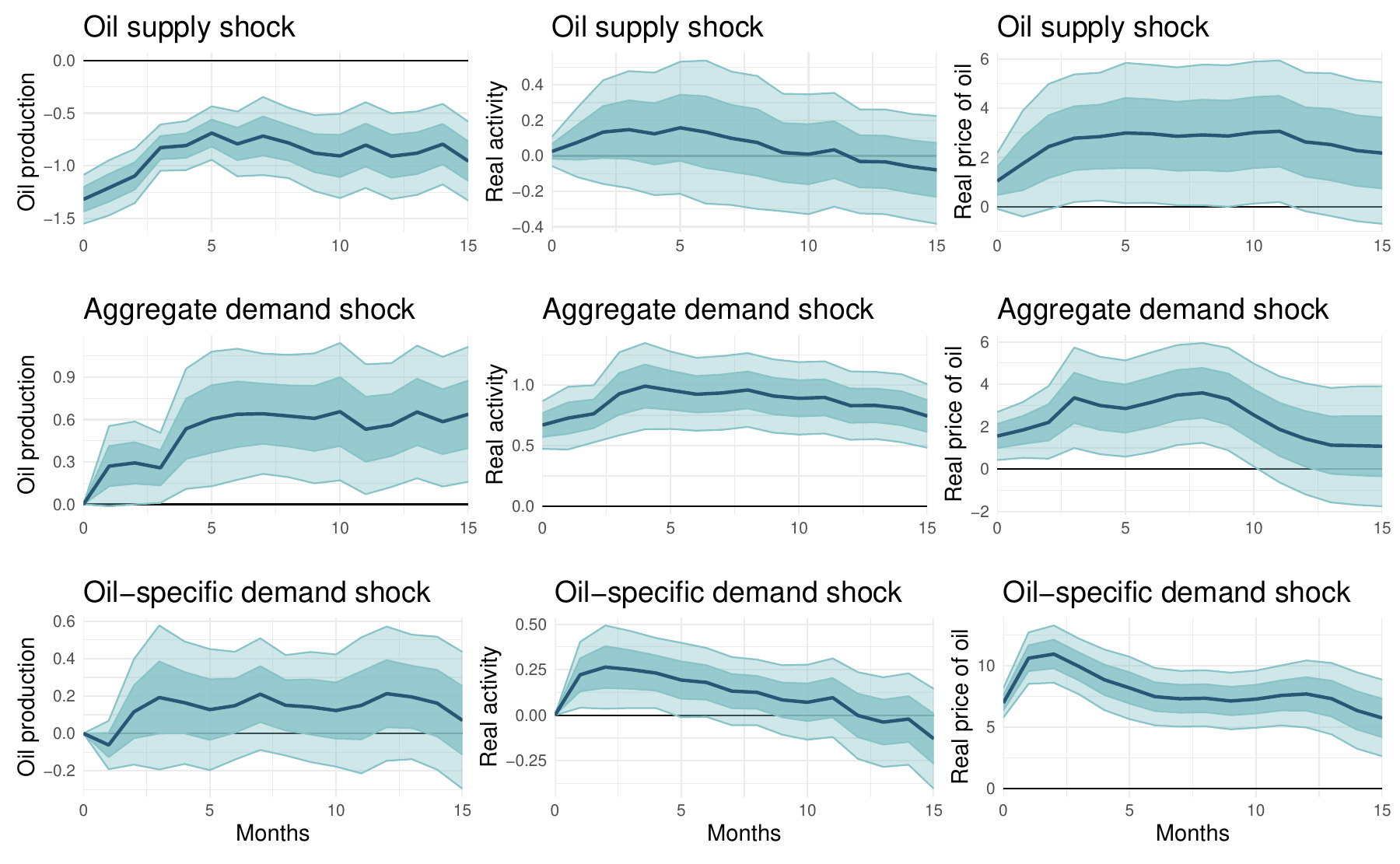}}
\caption[]{\label{fig:irf-wip} Responses of one-standard-deviation structural shocks (based on an index of industrial production)}
\begin{figurenotes}[Note]
  Solid lines show the point estimates and the shaded regions show one- and two-standard-error bands.
  Estimates are based on the structural VAR model in \eqref{eq:svar},  
  using the updated data (\dateFullWIPStart--\dateFullWIPEnd).
  Confidence intervals were constructed using a residual block bootstrap procedure \citep{bruggemann_jentsch_trenkler_2016}.  
\end{figurenotes}
\end{figure}

The structural VAR representation of this model is the same as in \eqref{eq:svar}
with $y_{t}$ replaced by 
$\tilde{y}_t = \left(\right.\Delta \text{oil production}$,
  $\Delta \text{OECD+6 industrial production}$, $\text{real price of oil} \left.\right)^{\prime}$.
The model is estimated using the same timing assumptions as the ones specified in \eqref{eq:identification}.
Figure \ref{fig:irf-wip} shows the structural impulse response functions, which are
the responses of elements of $\tilde{y}_{t}$ to one-time impulses in structural innovations. 
Because industrial production is measured as the difference of logs,
all responses for this variable are summed to approximate the cumulative percent change.
Again,
the innovations are normalized so that each innovation will tend to raise the price of oil.

Figure \ref{fig:irf-wip}, which is based on OECD+6 industrial production,
is comparable to figure \ref{fig:irf-full}.
In brief,
the broad patterns are remarkably similar.
Our results are in line with the results reported by \citet{zhou_2020}.

\subsubsection{Historical decomposition}

\begin{figure}[htbp]
\centerline{\includegraphics[width=\textwidth]{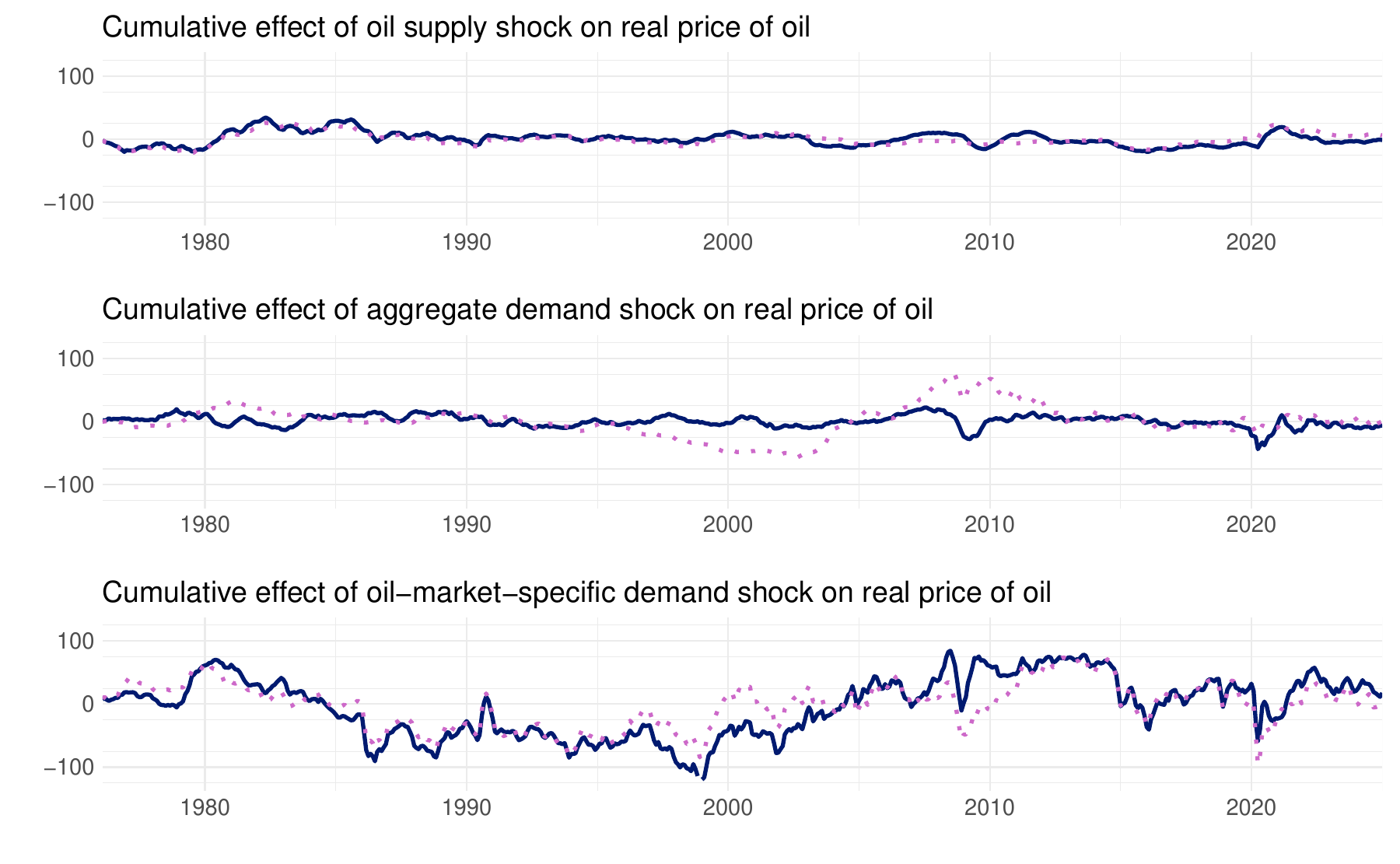}}
\caption[]{\label{fig:hist-decomp-wip} Historical decomposition of the real price of oil}
\begin{figurenotes}[Note]
  \textit{Top}, how much oil-supply shocks explain of the historically observed fluctuations in the real price of crude oil;
  \textit{middle}, how much aggregate-demand shocks explain of the historically observed fluctuations in the real price of crude oil;
  \textit{bottom}, how much oil-specific-demand shocks explain of the historically observed fluctuations in the real price of crude oil.
  Estimates are based on the structural VAR model in \eqref{eq:svar},
  using data from \dateFullWIPStart{} to \dateFullWIPEnd. 
  The solid, blue lines are based on the growth rate of industrial production
  instead of using the index of global real economic activity described in the main text \citep{kilian_2009,kilian_2019}.
  The dotted, purple lines correspond to statistics from figure \ref{fig:hist-decomp},
  which are based on \citeauthor{kilian_2009}'s \citeyearpar{kilian_2009} index of global real economic activity.
\end{figurenotes}
\end{figure}

Given that the impulse response functions are very similar,
it is not surprising that the historical decomposition is similar as well.
Figure \ref{fig:hist-decomp-wip} shows the cumulative contribution of
oil-supply, aggregate-demand, and oil-specific-demand shocks to the real price of oil in separate panels.
The contributions based on the industrial-production index are shown using solid, blue lines.
The contributions shown with dotted, purple lines are based on \citeauthor{kilian_2009}'s \citeyearpar{kilian_2009} index.
(These are copied from figure \ref{fig:hist-decomp}.)

The cumulative contributions---whether based on \citeauthor{kilian_2009}'s \citeyearpar{kilian_2009} index or the index of industrial production---are broadly similar.
Some differences are expected, however.
As \citet{zhou_2020} explains,
the timing of the two indicators of global real economic activity differ. 
\citeauthor{kilian_2009}'s \citeyearpar{kilian_2009} index is ``a leading indicator with respect to global real output;''
whereas,
the industrial-production index is a coincident indicator of global real output and therefore
``[misidentifies] the timing of shifts in aggregate demand in commodity markets'' \citep[][58]{kilian_zhou_2018}.
Regardless,
insights made by \citet{kilian_2009} continue to hold:
oil-supply shocks have made small contributions to the real price of oil and
oil-specific-demand shocks have contributed the ups and downs to the real price of oil.

\subsubsection{Responses in Kern County, California, to the global oil market}

We also investigated the robustness of the second-stage regressions that investigate the responses of labor-market outcomes in Kern County, California.
The patterns based on shocks estimated using industrial production are similar to those shown in figure \ref{fig:irf2-kern}.
We therefore skip reporting the results here and, instead,
report the figure that corresponds to figure \ref{fig:irf2-kern} in our replication materials.

\end{document}